\documentclass{INTERSPEECH2023}

\interspeechcameraready

\title{A Generative Framework for Conversational Laughter:
  Its `Language Model' and Laughter Sound Synthesis}

\name{Hiroki Mori$^1$, Shunya Kimura$^1$}
\address{
  $^1$Utsunomiya University, Japan}
\email{hiroki@speech-lab.org}

\begin{document}

\maketitle
 
\begin{abstract}
  %% Laughter synthesis is an emerging technology, and gaining its
  %% importance as the human-agent interaction gets more advanced and
  %% more popular in our life. However, the use, phonetic structure,
  %% and functional aspects of laughter in conversation
  %% has not been well explored, which has prevented to set a rigid
  %% research goal of laughter synthesis. 
As the phonetic and acoustic manifestations of laughter in conversation are highly diverse, laughter synthesis should be capable of accommodating such diversity while maintaining high controllability. This paper proposes a generative model of laughter in conversation that can produce a wide variety of laughter by utilizing the emotion dimension as a conversational context. The model comprises two parts: the laughter ``phones generator,'' which generates various, but realistic, combinations of laughter components for a given speaker ID and emotional state, and the laughter ``sound synthesizer,'' which receives the laughter phone sequence and produces acoustic features that reflect the speaker's individuality and emotional state. The results of a listening experiment indicated that conditioning both the phones generator and the sound synthesizer on emotion dimensions resulted in the most effective control of the perceived emotion in synthesized laughter.
\end{abstract}
\noindent\textbf{Index Terms}: laughter synthesis, generative model,
language model of laughter, emotional conditioning

\section{Introduction}

Laughing is a basic and essential emotional behavior for humans.
Nevertheless, almost all of the conversational agents that interact
with humans do not laugh.
Part of the reason for this is attributed to the fact that we ourselves
do not well understand why, when, and how we laugh.
A recent study on conversational robots by a Kyoto-U team
aimed at the positive effect
of the robot's laughter on empathy \cite{Inoue2022}.
By focusing on ``shared laughter,'' they cleared the \emph{when} problem.
For the \emph{how} problem, however, they avoided laughter synthesis
and randomly picked one from the pools of ``mirthful'' or
``social'' laughs.

%% However, the use, phonetic structure, and functional aspects of
%% laughter in conversation has not been well explored, which has
%% prevented to set a rigid research goal of laughter synthesis.

The current laughter synthesis
study focuses on \emph{how} conversational agents should
laugh. 
Laughter synthesis is an emerging technology and is gaining
importance as the human-agent interaction becomes more advanced and
popular in our daily lives
\cite{Trouvain2004,Urbain2014,mori2019_is,nagata2020,tits2020,arimoto2022}.
A large part of previous work has employed a similar framework to
text-to-speech systems. An open problem here is
how to construct input sequences for the synthesizer.
As the phonetic structure and its functional aspects of laughter
in conversation has not been fully understood,
most previous work simply used exemplars of natural laughter
for input, which limits flexibility.
Recent research in non-speech vocalization synthesis \cite{hsu2022}
also points to the need for some kind of ``language model''.

This paper proposes a generative model of laughter in conversation
that can produce a wide variety of laughter.
A highlighted feature is the ``language model'' of laughter,
which serves as a laughter sequence generator.
This model generates various but
realistic combinations of laughter components for a given speaker ID
and emotional state.
The generated sequence is then fed into the laughter
``sound synthesizer,'' which produces acoustic features
that reflect the speaker's individuality and emotional state.

In this paper, we will be using specific laughter-related terminology,
following to \cite{Trouvain2003}.
A ``laughter episode'' will refer to
a series of acoustic events that correspond to exhalation or
inhalation. A ``bout'' will refer to an event that
corresponds to an exhalation and is composed of one or more laughter
calls. A ``call'' will be used to describe an individual unit of
laughter, analogous to a syllable. Therefore, a typical bout
``hahaha'' is a 3-call bout.

%\section{Collection and annotation of laughter sounds}
\section{Morphology of laughter sounds}
\label{sec:morph}
A typical method for collecting laughter data has been induction
by funny movies \cite{Bachorowski2001a,Urbain2010}.
Provine criticized past studies for focusing solely on
audience-oriented, passive laughter \cite{Provine2001}.
He argued that laughter is social and that speakers actually
laugh more than listeners.
Since we are interested in laughter in agent-human interaction,
we need to
collect laughter that occurs naturally in conversation.
In this study, we used the Online Gaming Voice chat
Corpus (OGVC) \cite{Arimoto2012a}, a speech corpus containing
spontaneous dialogue
during massively multiplayer online
role-playing games (MMORPGs), which has a larger number of laughs
than other Japanese conversational corpora used in emotion
studies.

\begin{figure*}[t]
\centering
  \includegraphics[width=.85\hsize]{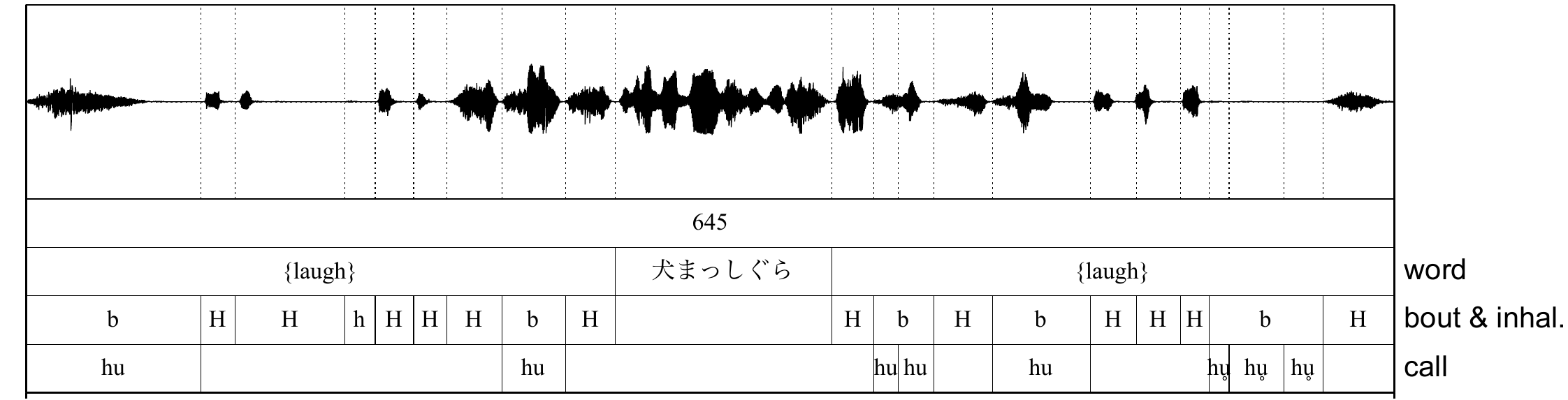}
  \caption{Bout- and call-level annotation of laughter.}
  \label{fig:06_FWA_645}
  \vspace*{-2mm}
\end{figure*}

Bout- and call-level annotation was performed for the top three
speakers with the highest frequency of laughter in OGVC.
An example of the annotation is shown in
Fig.~\ref{fig:06_FWA_645}.
The annotation has a hierarchical structure:
Bouts and inhalation sounds that comprise each laughter episode
were annotated, as well as calls that comprise each bout.

\begin{figure}[tb]
  \centering
  \includegraphics[width=\hsize]{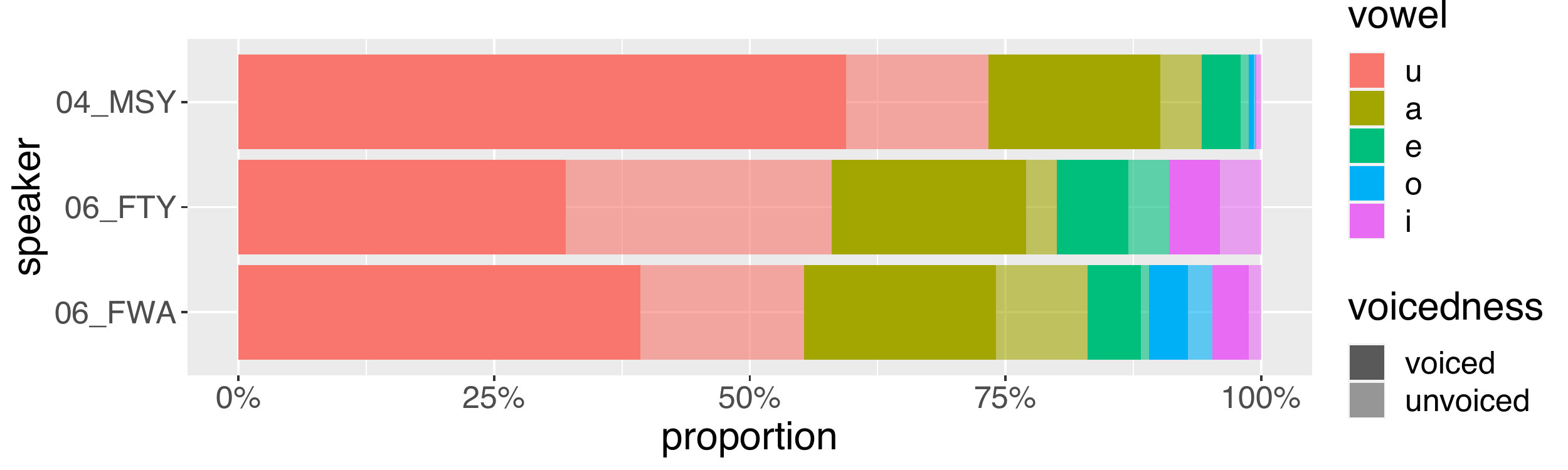}
  \caption{Vowel proportions of calls. Each darker color stands for
voiced, and lighter color for unvoiced.}
  \label{fig:vowel}
  \vspace*{-1mm}
\end{figure}

The consonant and vowel of each call were transcribed as a romanization of
Japanese syllable, rather than in a phonetic way. Therefore,
laughter vowels are classified into one of a, e, i, u, or o.
The proportions of vowels are shown in Fig.~\ref{fig:vowel}.
The most common vowel was /u/, followed by /a/.
However, these are not contrastive, and most laughter sounds
are realized around the mid central vowel [\textschwa].

In addition to consonants and vowels, phonetic variants, including
unvoiced (e.g.\ h\textsubring{u}), nasal (e.g.\ h\~u), and consonant
prolongation (e.g.\ h{\textlengthmark}u), were also transcribed.
Among them, the voicelessness of laughter sound has received much
attention due to its functional importance.
For example, voiced laughter induces significantly more
positive emotional responses in listeners than unvoiced laughter does
\cite{Bachorowski2001b}.

\begin{figure}[tb]
  \centering
  \includegraphics[width=\hsize]{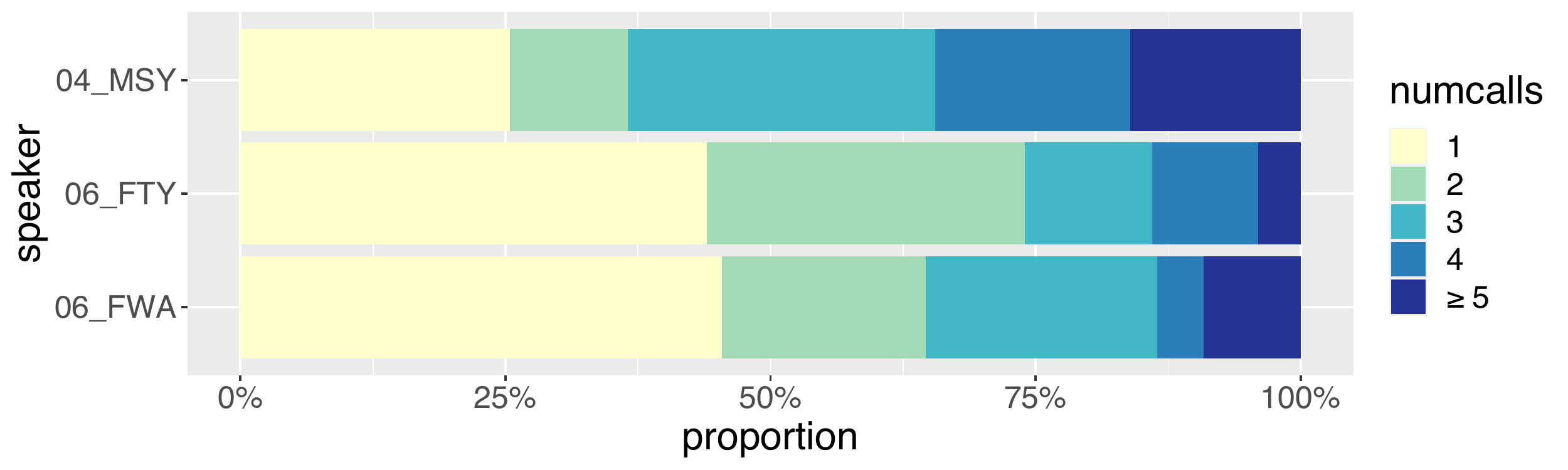}
  \caption{Proportions of the number of calls per bout.}
  \label{fig:numcalls}
\end{figure}

The proportion of bout length (number of calls)
is shown in Fig.~\ref{fig:numcalls}.
It is worth noting that the proportion of single-call bouts is
surprisingly large. The proportion of unvoiced calls in single-call
bouts (55.4~\%) is significantly larger than that of multi-call bouts
(18.7~\%). This suggests that single-call bouts tend to be
accompanied by negative emotions \cite{Bachorowski2001b}.

Individual inhalation sounds were identified as h (unvoiced) or H
(voiced), and annotated at the same tier as bouts. Inhalation
sounds often accompany vocal fold vibration (voiced), some of which
constitute a main part of a laughter sound. This voiced/unvoiced
distinction is crucial because of its relation to perceived emotion.
Arimoto et
al.\ \cite{arimoto2022} showed that laughs containing voiced inhalation
sounds tend to be perceived as more pleasant and aroused.
Voiced inhalation sounds are also important in characterizing the
individuality of laughing speakers.
For the top seven OGVC
speakers with the highest frequency of laughter,
the proportion of episodes with mid-laugh voiced inhalations is less
than 1~\% for two speakers, around 10~\% for three speakers, and 21~\% and
27~\% for the remaining two speakers. This implies that there are speakers
who almost exclusively use egressive laughter, as well as those who
frequently produce ingressive laughter.

\section{Emotion perception from laughter}
\label{sec:emotion}
The morphological variation of laughter depends on its discourse and
social context. However, it is difficult to encode such contexts in a
comprehensive and adequate way. As a first-order approximation,
this study attempts to use the
speaker's emotion perceived from laughter as an explanatory variable
for modeling laughter forms \cite{arimoto2022}.

This requires an evaluation of the perceived emotion for the laughs in
the corpus.
For this purpose, emotion categories such as ``big six'' emotions
\cite{Ekman1999a} seem virtually useless.
In this study, we annotated the emotion perceived from laughter
with two emotion dimensions, pleasantness and arousal.
Dimensional descriptions of emotions
have a long history and are well established in psychology.
A number of studies have stated that two or three dimensions are sufficient
to account for a good portion of emotional variation. Among all,
the pleasantness (also known as valence) and arousal (also known as
activation) dimensions have been regarded as fundamental \cite{russell1997}.

Prior to the emotion annotation, the first author checked the laughter
sounds of a male speaker 04\_MSY and a female speaker 06\_FWA, then
filtered out subtle or less audible ones, which yielded 125 and 100
laughter episodes for the two speakers as our laughter dataset.

\begin{figure}[tb]
  \centering
  \includegraphics[width=.8\hsize]{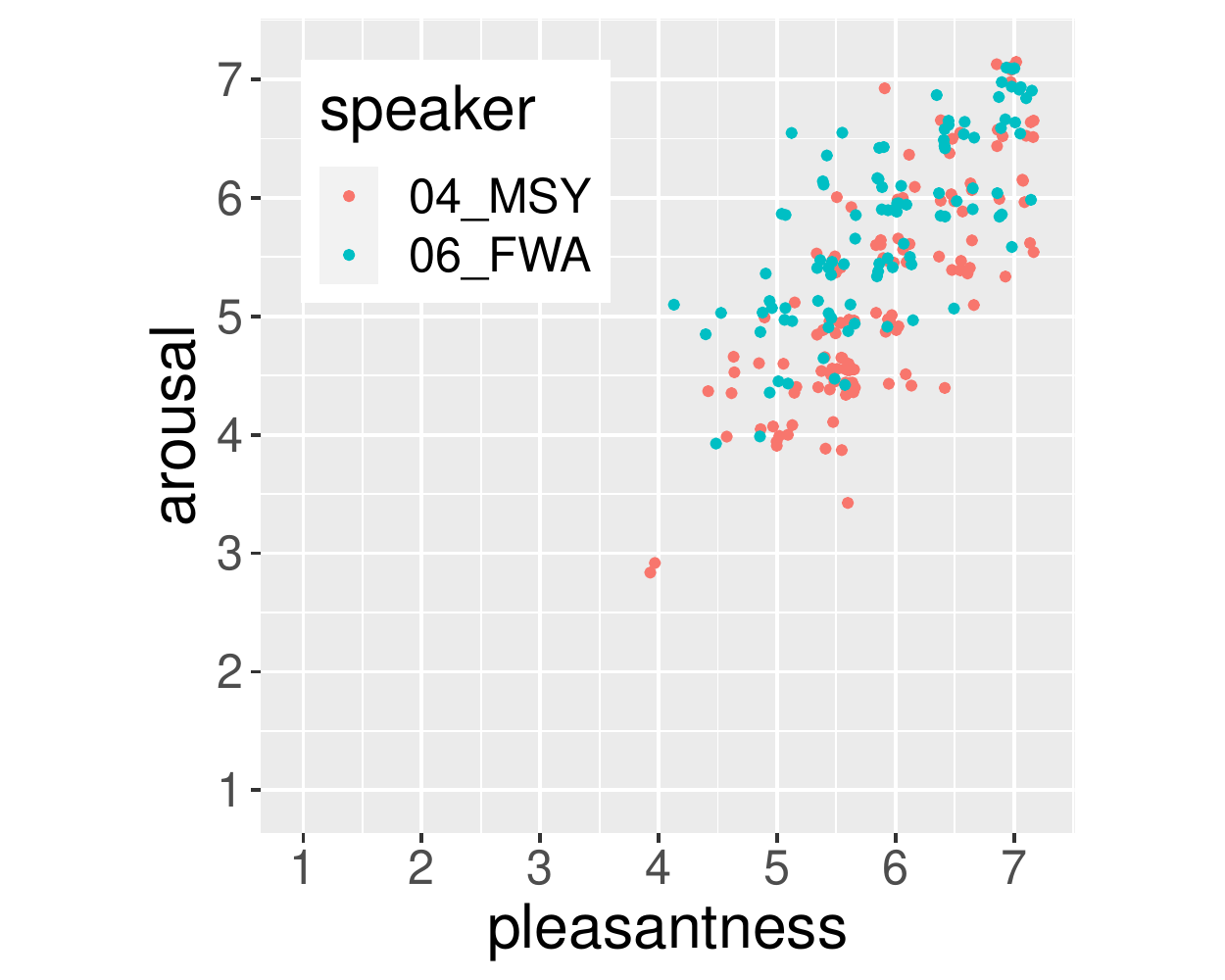}
  \caption{Distribution of the ground-truth pleasantness and arousal
dimensions evaluated for laughter sounds. Points are jittered to avoid
overplotting of laughters with identical values.}
  \label{fig:emodim-laughter}
\end{figure}

The two authors individually annotated the perceived pleasantness (1:
extremely unpleasant, 7: extremely pleasant) and arousal
(1: extremely sleepy, 7: extremely aroused). The
ground-truth values were obtained by averaging them.
Figure~\ref{fig:emodim-laughter} shows the distribution of the emotion
dimensions for the two speakers.
Most laughter sounds were evaluated as more pleasant and more aroused
than neutral (4). Mean pleasantness and arousal were 5.86 and 5.19 for
the male speaker 04\_MSY, and 5.95 and 5.78 for the female speaker 06\_FWA.

\section{Phones generator: The ``language model'' of laughter}
Contrary to the notion that laughter sounds have a homogenious
structure such as ``hahaha,'' ``hehehe,'' or ``huhuhu,''
there are so many variations that
a closed lexicon of laughter cannot be defined.
 At the same time, we barely hear
laughter sounds such like ``hahohaho,'' which implies that there are
some constraints that prescribe possible combinations of laughter calls.
Provine \cite{Provine2001} suggested biological constraints
against producing such mixed-call laughs, but he also pointed out
that one can easily switch call types in mid-laugh, as in
``hahahoho.'' His observation implied the existence of some
laughter \emph{grammar}, but he did not discuss a computational model
of laughter calls that could be applied to laughter synthesis.
%% Moreover, he did not argue the inhalation sounds that associate
%% laughter bouts, which the authors think are essential components of
%% conversational laughter.

A desired laughter language model should not only regulate such possible
combinations (as opposed to the random arrangement \cite{arimoto2022}),
but also account for morphological preferences related to
discourse and social context.
As described in Sect.~\ref{sec:morph}, the length of laughter is related to
its emotion. Therefore, we modeled the length first, then
the components. Hereafter, we regard either a call or a single inhalation
as a component and refer to each component as a ``phone.''
For example, the phone sequence corresponding to the second laughter episode in
Fig.~\ref{fig:06_FWA_645} is  ``H hu hu H hu H H H
h\textsubring{u} h\textsubring{u} h\textsubring{u} H.''

\begin{figure}[tb]
  \centering
  \includegraphics[width=\hsize]{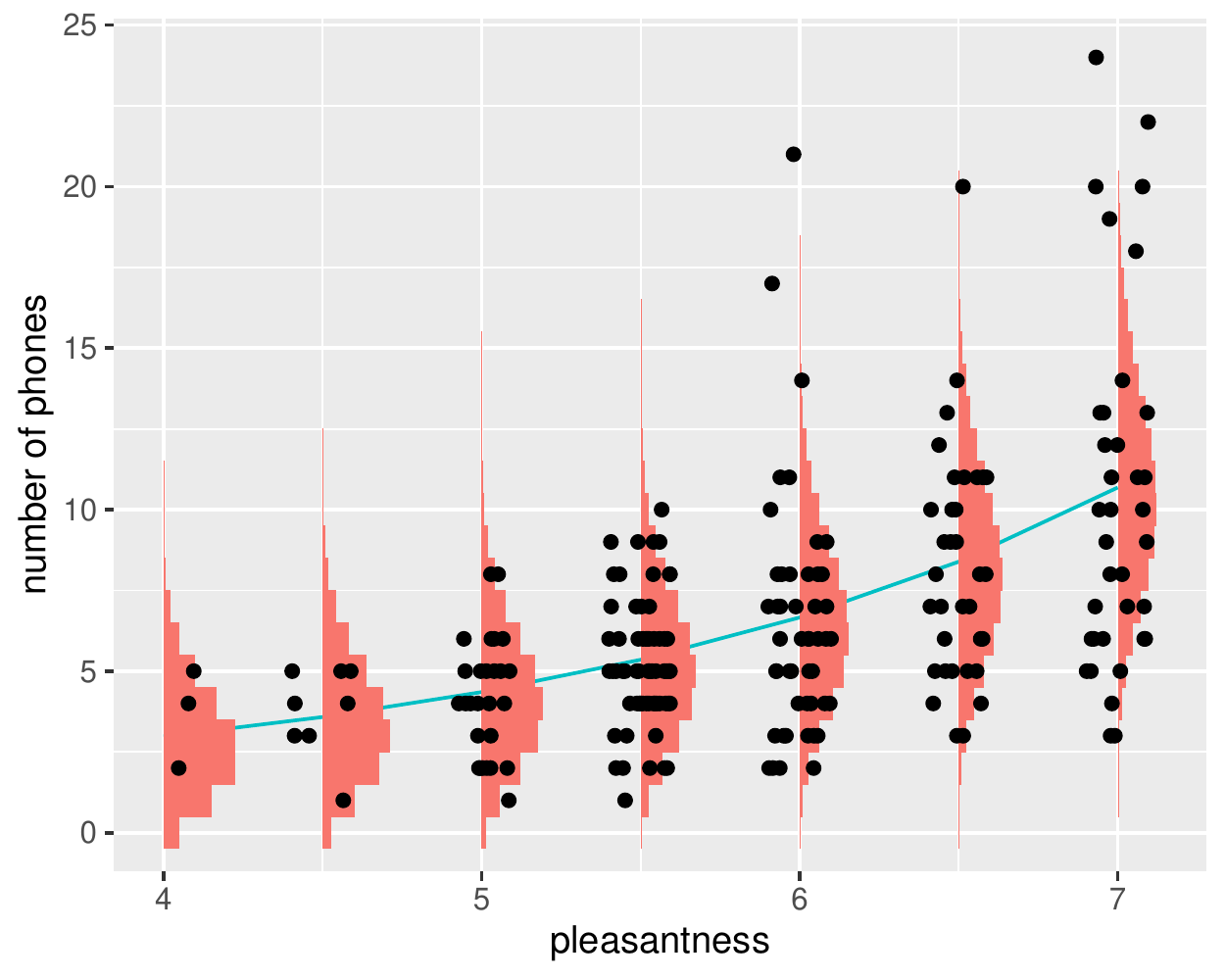}
  \caption{Laugh length distribution (points) and its probabilistic model with
Poisson regression. Points are horizontally jittered to avoid
overplotting.}
  \label{fig:poisson}
\end{figure}

Figure~\ref{fig:poisson} shows the
distribution of the laugh length (number of phones)
versus pleasantness by black points.
As these could be
modeled by a Poisson regression, the fitted mean parameter $\lambda$ (green
line) and probability mass function (red bars) are overlaid
(here the arousal value was set equal to the pleasantness value for simplicity).
A generalized linear model with Poisson distribution was obtained
through variable selection using AIC (Akaike Information Criterion) \cite{james2013ch6}.
The fitted laugh length model was as below:
\begin{align}
  \log(\lambda_i)&=b+0.527x_i^\text{ple}+0.750x_i^\text{ple}x_i^\text{aro}, \label{eq:glm}
  \\
  y_i &\sim \text{Pois}(\lambda_i),
\end{align}
where \(y_i\) is the length of $i$-th laughter,
\(x_i^\text{ple}\) and \(x_i^\text{aro}\) are the pleasantness and
arousal dimensions, whose
range is linearly transformed from \([1,7]\) to \([-1,1]\),
and $b$ is the speaker specific baseline (1.433 for 04\_MSY,
0.936 for 06\_FWA).

In the generation phase, the decision to stop generating is determined
dynamically and randomly. Here we define
\(P_\text{end}(n)\) as the probability that the $n$-th generated phone
is the last one:
\begin{equation}
  P_\text{end}(n)=\frac{f(n;\lambda)}{1-F(n-1;\lambda)},
\end{equation}
where \(f(k;\lambda)\) and \(F(k;\lambda)\) are
the probability mass function and cumulative distribution function
of \(\text{Pois}(\lambda)\), respectively.
For each generated phone, an ``end-of-laughter''
is drawn according to \(P_\text{end}(n)\). This ensures that the length
distribution of generated laughs follows the Poisson distribution,
whose mean is determined by Eq.~(\ref{eq:glm}).

%The ``vocabulary'' of laughter component was built according to the
%call transcription plus unvoiced/voiced inhalation (h/H).
Thirty-two different phones appeared in the laughter dataset described in
Sect.~\ref{sec:emotion}. By replacing phones that appeared
only once (e.g.\ h{\textlengthmark}a, h\textsubring{i}, na) with
similar ones, we obtained a phone list comprising 22 different
calls and inhalations. In the modeling, phone sequences that constitute
each laughter episode were converted into a sequence of 64-dimensional
embedding vectors.

Similar to neural language models \cite{merity2018},
the call sequence of laughter was modeled with a recurrent neural
network.
We used an architecture with
an LSTM layer with 128 hidden dimensions, a linear layer, and a
softmax layer.
The dimensionality of the input was 64 (phone embedding)
\({}+{}\) 1 (\(P_\text{end}(n)\))
 \({}+{}\) 1 (speaker)
 \({}+{}\) 2 (emotion dimensions) \(=68\).

\begin{figure}[tb]
  \centering
  \includegraphics[width=.8\hsize]{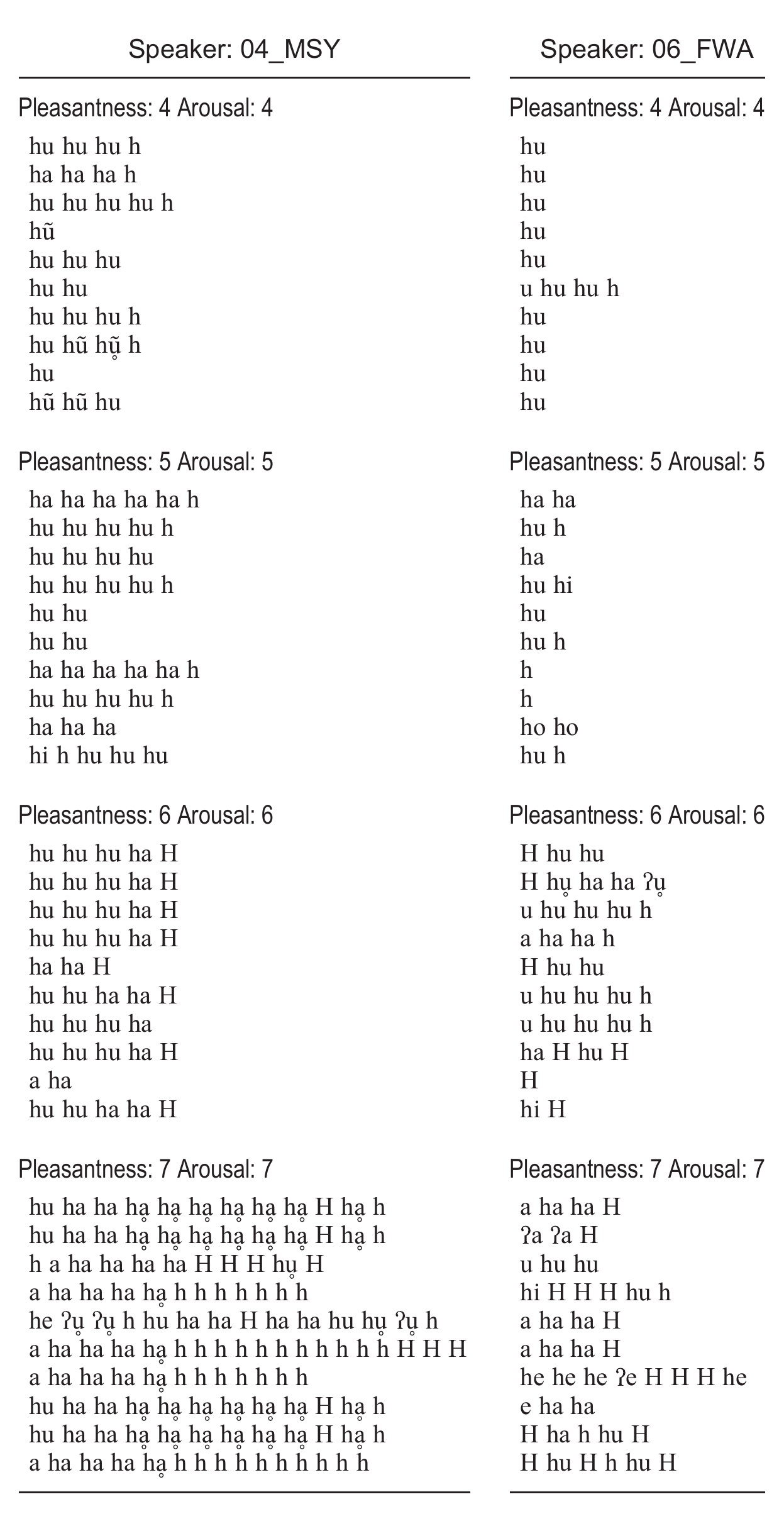}
  \caption{An excerpt from generated laughter phones (10 draws per
    condition).
  See the multimedia file for the complete list.}
  \label{fig:calls}
\end{figure}

Generated phones resulting from 10 draws for several combinations of emotion
dimensions are shown in Fig.~\ref{fig:calls}.
Note that these are random draws without any cherry-picking, so many
duplicates exist in the lists.
From the figure, it is apparent that emotion and speaker individuality
are reflected not only in the length of laughter
but also in the pattern of laughter phones. For example, more pleasant and
aroused laughter contains more /a/'s and voiced inhalations.

\section{Laughter sound synthesizer}
The current waveform synthesizer is basically a vocoder-based parametric
speech synthesis \cite{zen2013}, which can model human
vocalization better than end-to-end models for limited data sizes,
such as the one used in our case. The input feature set for duration modeling
consisted of the identity of current consonant-vowel (19),
2 phonetic variations
(voicedness, nasality) and their left and right context (\texttimes3),
phone position (1),
laughter length (1), and 2 emotion dimensions (67 in total).
For acoustic modeling, 
phone duration and 3 numerical features for coarse-coded frame
position in the current phone \cite{zen2015} were added to the input,
and 59th-order Mel-cepstrum, \(\log f_o\),
aperiodicity, their $\Delta$, $\Delta\Delta$, and the voicedness
were inferred as the output.
The network
was composed of a three-layer stacked bidirectional LSTM with 128 hidden
dimensions and a linear layer. For the subsequent experiment,
the model was trained with the 04\_MSY dataset whose waveform was downsampled
to 16~kHz.

\section{Experiment}
To investigate emotion controllability in the proposed
laughter synthesis, we conducted an ablation study on both the
phones generator
and the laughter sound synthesizer.
Hereafter, we denote the absence or presence of emotion inputs to the
phones generator as \textminus/+phones, and similarly,
the absence or presence of emotion inputs to the laughter sound
synthesizer as \textminus/+acoust.
The emotion inputs were masked at the training and
inference stages in the {\textminus}phones and {\textminus}acoust conditions.
For each of 10 pleasantness and arousal combinations
(4, 4), (4, 5), (5, 4), (5, 5), (5, 6), (6, 5), (6, 6), (6, 7), (7,
6), and (7, 7), twenty sequences were generated using the
phones generator. Then, the corresponding laughter waveform was
synthesized from the acoustic features generated for each sequences
using WORLD \cite{Morise2016}. The generated phones and synthesized
waveforms are provided as the multimedia files for this paper.

For each condition, the first 10 phone sequences were used in the
listening test (see Fig.~\ref{fig:calls}).
The number of stimuli was 10 (target emotion dimensions) ${}\times{}$
10 (phone sequences)  ${}\times{}$
2 (\textminus/+phones) ${}\times{}$
2 (\textminus/+acoust) plus two reference real laughter
sounds for subject screening ${}\times{}$ 4 (repetitions) ${}=408$.
Thirty-one undergraduate and graduate students who were not involved in
speech research participated in the listening test.
First, they watched a video that described the objectives of the
experiment and an introduction to the theory of emotion dimensions.
The subjects then used a web interface
to listen to the stimulus sounds in a random order
and evaluated perceived pleasantness and arousal on a 7-point scale,
as in Sect.~\ref{sec:emotion}.
From the results of the screening test,
two subjects were found not to meet our criteria (distinguishing between
obviously pleasant/unpleasant laughter and responding consistently to
identical stimuli), so their responses were excluded from later analysis.

\begin{figure}[tb]
  \centering
  \includegraphics[width=\hsize]{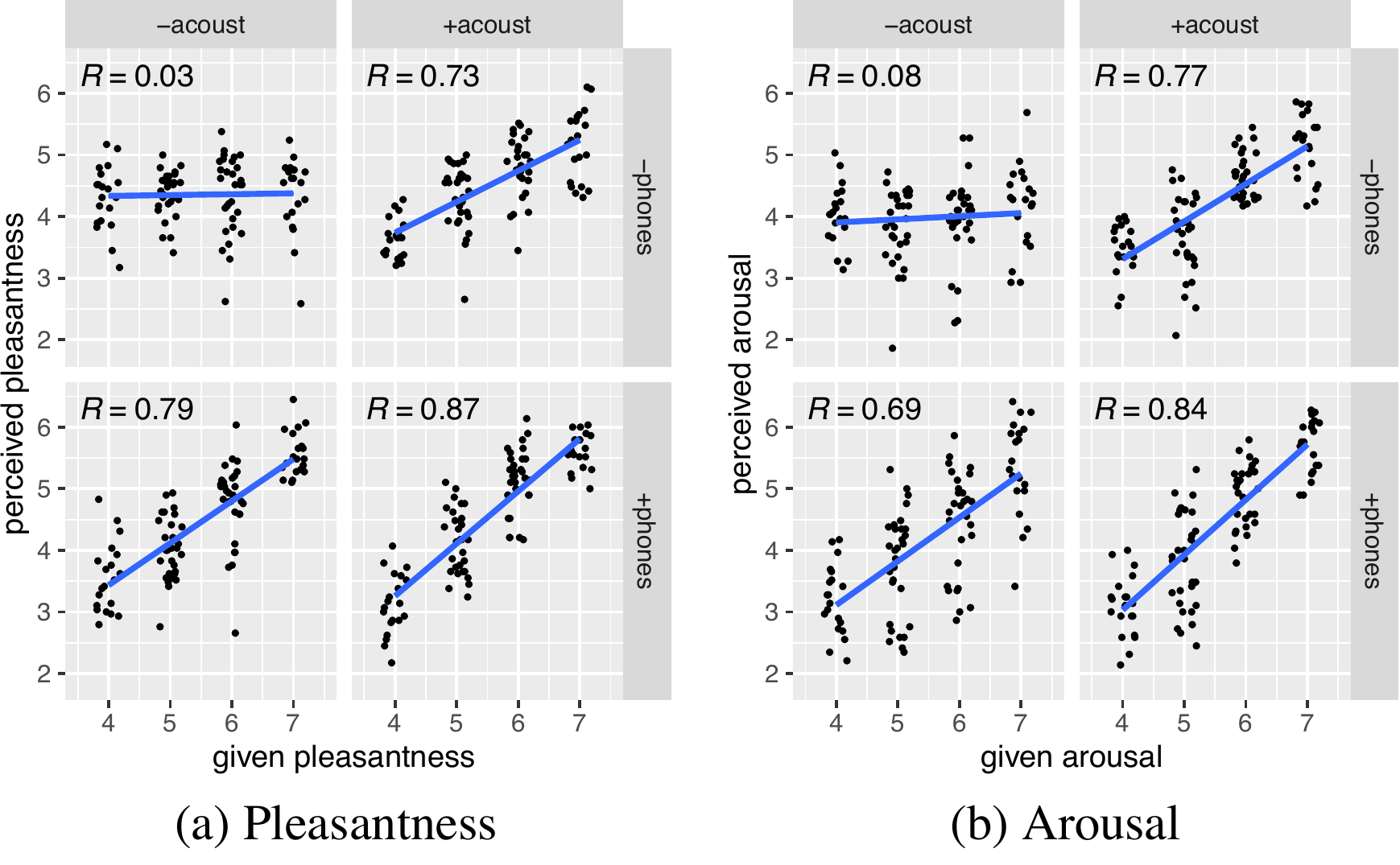}
  \caption{Relationship between target and perceived emotion from
    synthesized laughter for (a) pleasantness, and (b) arousal.
 Points are horizontally jittered to
avoid overplotting.}
  \label{fig:correlation-bind}
\end{figure}

The perceived pleasantness and arousal for the 400 synthesized laughter sounds were averaged
over the subjects. Figure~\ref{fig:correlation-bind}
shows the distribution of perceived
pleasantness and arousal.
For both dimensions, the +phones+acoust model showed the best
controllability, as the correlation coefficient is as high as
0.87 (pleasantness) and 0.84 (arousal).
This means that the emotion input to the phones generator
and the emotion input to the laughter sound synthesizer
are individually effective, but the emotion input to the both modules
is even more effective.
A statistical test for the difference
between two paired correlations revealed that the correlation
coefficient for the +phones+acoust model is significantly higher
than that for the +phones{\textminus}acoust model for both dimensions
(\(p<0.01\)).

Best linear models to predict responses from the target dimension were
obtained through variable selection using AIC:
\begin{align}
  \hat{y}^\text{ple}=& 0.112-0.301\delta_\text{phones}-0.199\delta_\text{acoust}
  +0.141\delta_\text{phones}\delta_\text{acoust} \nonumber \\
  &{}+(0.669\delta_\text{phones}+0.489\delta_\text{acoust}-0.321\delta_\text{phones}\delta_\text{acoust})x^\text{ple},
\label{eq:yple}\\
  \hat{y}^\text{aro}=& -0.265\delta_\text{phones}-0.198\delta_\text{acoust}
  +0.171\delta_\text{phones}\delta_\text{acoust} \nonumber \\
  &{}+(0.661\delta_\text{phones}+0.561\delta_\text{acoust}-0.375\delta_\text{phones}\delta_\text{acoust})x^\text{aro}, \label{eq:yaro}
\end{align}
where \(\delta_\text{phones}\) and \(\delta_\text{acoust}\) are the
dummy (0/1) variables corresponding to the \textminus/+phones and
\textminus/+acoust
conditions.
The coefficients of \(x^\text{ple}\) and \(x^\text{aro}\)
in Eqs.~(\ref{eq:yple}) and (\ref{eq:yaro}) clearly
demonstrate the synergistic effect gained by controlling both
the phones generator and the sound synthesizer.

\section{Conclusions}

In this paper, we proposed a generative model for laughter in conversation,
which allows for the production of
a wide variety of laughter that can be controlled by
emotion dimensions. Our results indicate that
conditioning both the phones generator and the
laughter sound synthesizer on emotion dimensions is most effective
in controlling perceived pleasantness (\(R=0.87\))
and arousal (\(R=0.84\)).

One limitation of the current study is the lack of scalability,
as call-level annotation for new datasets could become a bottleneck.
Although state-of-the-art speech recognition systems such as Whisper
can transcribe laughter calls to some extent,
they cannot distinguish the phonetic variants necessary for laughter
synthesis.
One potential solution is to fine-tune the model using richly
annotated laughter data such as the one built in this study.

\section{Acknowledgements}

This work was supported by
\ifinterspeechfinal
JSPS KAKENHI Grant Numbers 22K12107 and 22K18477.
\else
xxxx.
\fi

\bibliographystyle{IEEEtran}
\bibliography{myreference}

\end{document}